\documentclass[a4paper,superscriptaddress,amsmath,amssymby,longbibliography]{revtex4-1}
\usepackage[english]{babel}
\PassOptionsToPackage{english}{babel}

\usepackage[utf8]{inputenc}
\usepackage{color}
\usepackage{amsmath,amssymb,graphicx,xcolor,units,hyperref}
\usepackage{booktabs}
\usepackage{soul}

\newcommand{\bra}[1]{\langle{#1}\rvert}
\newcommand{\ket}[1]{\lvert{#1}\rangle}

\newcommand{\ie}{\textit{i.\,e.}}

\newcommand{\micron}{\mu\mathrm{m}}

\begin{document}
\selectlanguage{english}
\title{Tunable photonic heat transport in a quantum heat valve}

\author{Alberto Ronzani}
\email{alberto.ronzani@aalto.fi}
\affiliation{QTF Centre of Excellence, Department of Applied Physics, Aalto
University School of Science, P.O. Box 13500, 00076 Aalto, Finland}

\author{Bayan Karimi}
\affiliation{QTF Centre of Excellence, Department of Applied Physics, Aalto
University School of Science, P.O. Box 13500, 00076 Aalto, Finland}

\author{Jorden Senior}
\affiliation{QTF Centre of Excellence, Department of Applied Physics, Aalto
University School of Science, P.O. Box 13500, 00076 Aalto, Finland}

\author{Yu-Cheng Chang}
\affiliation{QTF Centre of Excellence, Department of Applied Physics, Aalto
University School of Science, P.O. Box 13500, 00076 Aalto, Finland}
\affiliation{Department of Physics, National Taiwan University, Taipei, Taiwan,
Republic of China}
\affiliation{Institute of Physics, Academia Sinica, Taipei, Taiwan, Republic of China}

\author{Joonas T. Peltonen}
\affiliation{QTF Centre of Excellence, Department of Applied Physics, Aalto
University School of Science, P.O. Box 13500, 00076 Aalto, Finland}

\author{ChiiDong Chen}
\affiliation{QTF Centre of Excellence, Department of Applied Physics, Aalto
University School of Science, P.O. Box 13500, 00076 Aalto, Finland}
\affiliation{Institute of Physics, Academia Sinica, Taipei, Taiwan, Republic of China}

\author{Jukka P. Pekola}
\affiliation{QTF Centre of Excellence, Department of Applied Physics, Aalto
University School of Science, P.O. Box 13500, 00076 Aalto, Finland}

\begin{abstract}
Quantum thermodynamics is emerging both as a topic of fundamental research
and as means to understand and potentially improve the performance of quantum
devices~\cite{vinjanampathy_quantum_2016,martinezheat,goold_role_2016,pekola2015,jezouin,schwab,banerjee,sivre,cottet,partanen2017}.
A prominent platform for achieving the necessary manipulation of quantum states
is superconducting circuit quantum electrodynamics (QED)~\cite{wallraff}.
In this platform, thermalization of a quantum
system~\cite{neill_ergodic_2016,srednicki_chaos_1994,kaufman,reimann_eigenstate_2015}
can be achieved by interfacing the circuit QED subsystem
with a thermal reservoir of appropriate Hilbert dimensionality.
Here we study heat transport through an assembly consisting
of a superconducting qubit~\cite{koch_charge-insensitive_2007} capacitively
coupled between two nominally identical coplanar waveguide resonators, each
equipped with a heat reservoir in the form of a normal-metal mesoscopic resistor
termination.  
We report the observation of tunable photonic heat transport through the
resonator-qubit-resonator assembly, showing that the reservoir-to-reservoir heat
flux depends on the interplay between the qubit-resonator and the
resonator-reservoir couplings, yielding qualitatively dissimilar results in
different coupling regimes. 
Our quantum heat valve is relevant for the realisation of quantum heat
engines~\cite{rossnagel_single_2016} and refrigerators, that can be obtained, 
for example, by exploiting the time-domain dynamics and coherence of driven
superconducting qubits~\cite{kosloff,karimi_otto_2016}.  
This effort would ultimately bridge the gap between the fields of quantum
information and thermodynamics of mesoscopic systems.
\end{abstract}

\maketitle

Mesoscopic normal-metal (N) resistors are a natural candidate for the role
of heat reservoirs for superconducting circuit QED
experiments. Their geometry and transport properties can be adapted to provide a
controllable amount of dissipation by virtue of the electron-photon
interaction~\cite{schmidt_photon-mediated_2004,meschke,partanen2016}.
Furthermore, either clean or tunnel-type interfaces with the surrounding circuit
elements enable control of impedance mismatch for a given microwave design.  
With their fast internal thermalization timescales~\cite{pothier} and slow
electron-phonon relaxation~\cite{wallraff,simone} at subkelvin temperatures,
reservoirs formed of normal metal electrodes have been demonstrated as effective
broadband microwave detectors~\cite{govenius_microwave_2014} and
sources~\cite{tan_quantum-circuit_2017}. 
Their thermal properties, as well as the
experimental techniques required for temperature manipulation and readout, are
well established and understood~\cite{revmodgiaz}.

\begin{figure}
	\includegraphics{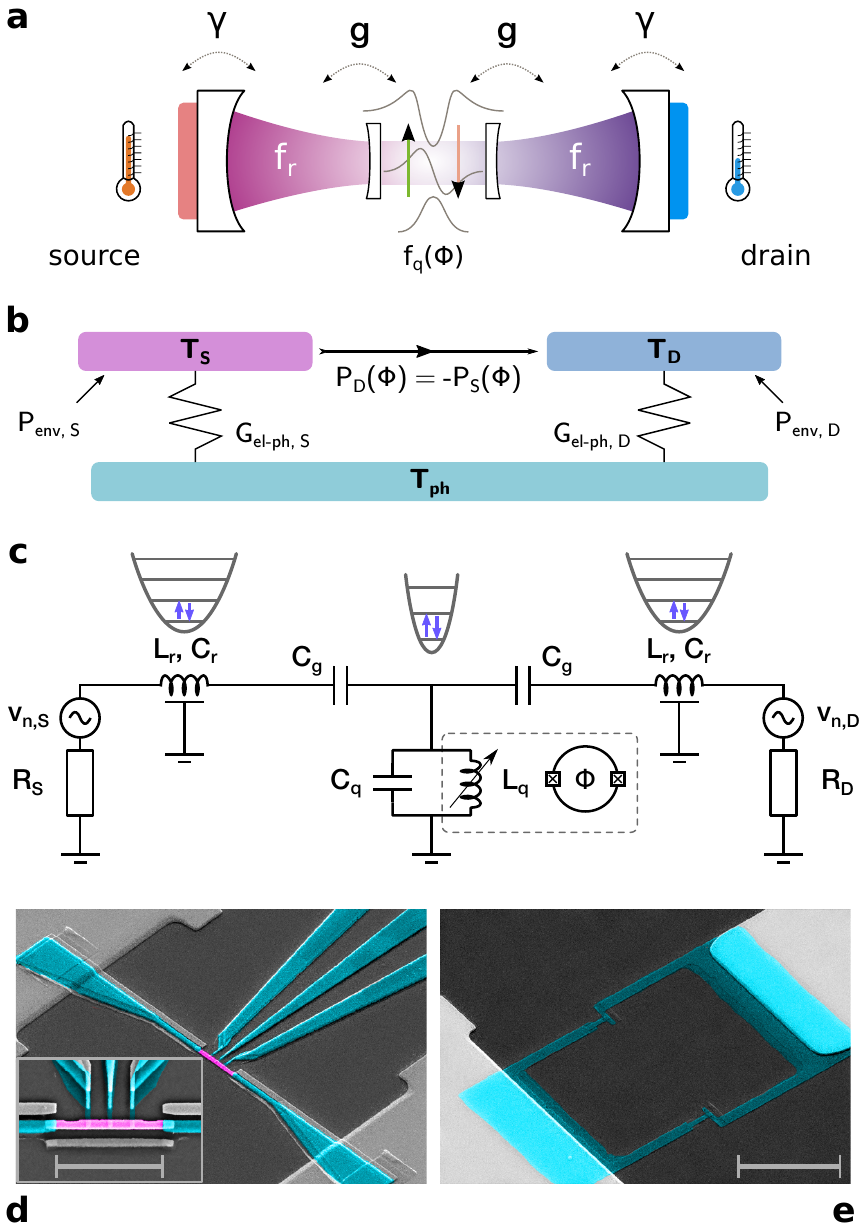}
	\caption{
	\textbf{Quantum Heat Valve design}.
	A transmon qubit having
	magnetic flux-tunable level spacing $f_\mathrm{q}(\Phi)$ is capacitively
	embedded between two superconducting transmission lines
	of identical length $= 4.6\,\mathrm{mm}$, each
	terminated by a mesoscopic normal-metal reservoir. We study the
	temperature of the drain reservoir $T_\mathrm{D}$ as a function of the
	temperature of the source reservoir $T_\mathrm{S}$ and of the ratio $r \equiv
	f_\mathrm{q}/f_\mathrm{r}$, where $f_\mathrm{r}$ is the fundamental resonant frequency of the
	transmission lines.  
	\textbf{a}: Conceptual depiction of the heat valve.
	\textbf{b}: Thermal model.
	\textbf{c}: Lumped-element idealisation of the device;
	capacitors $C_\mathrm{g}$ couple the transmon to each
	$L_\mathrm{r}C_\mathrm{r}$ resonator.
	\textbf{d}: Scanning electron micrograph of a waveguide
	termination, including three tunnel electrodes. 
	A copper resistor (pink, online) is in clean contact
	with aluminium leads (light blue, online) connecting to the patterned
	niobium film (light grey) on sapphire substrate (dark grey).
	The inset shows a magnified orthogonal view of the area spanned by the
	normal-metal element; the scale bar corresponds to $3\, \mathrm{\mu m}$.
	\textbf{e}: Scanning electron micrograph
	of the SQUID element in the transmon structure; the scale bar
	corresponds to $10\,\mathrm{\mu m}$.}	
	\label{fig:system}
\end{figure}

In this work we consider heat transmitted between two such mesoscopic
reservoirs, each of which tied to the photon occupation number of a microwave
resonator by the temperature-dependent Johnson-Nyquist current fluctuations of
the resistor.  Here, the two resonators are designed to have identical resonant
frequencies $f_\mathrm{r}$ and they are coupled to each other via a tunable
oscillator, a transmon-type qubit. This resonator-qubit-resonator assembly
constitutes a Quantum Heat Valve (QHV).
The thermal conductance of the QHV, conceptually depicted in
Figure~\ref{fig:system}a, is expected to depend on the reservoir-resonator and
resonator-qubit couplings (respectively $\gamma,\,g$, both normalised with
respect to $f_\mathrm{r}$) and on the ratio $r$ between the level spacing of the
qubit and the eigenfrequency of the resonators ($f_\mathrm{q} \equiv r f_\mathrm{r}$).  
For the transmon qubit, $f_\mathrm{q}$ depends on $\Phi$ as
\begin{equation}
	f_\mathrm{q}(\Phi) = \frac{\sqrt{8 E_\mathrm{J}(\Phi)
	E_\mathrm{C}}-E_\mathrm{C}}{h}\, ,
	\label{eqn:ftransmon}
\end{equation}
where $E_\mathrm{C}$ and $E_\mathrm{J}(\Phi) = E_\mathrm{J0} \left| \cos(\pi
\Phi/\Phi_0) \right| \sqrt{1 + d^2 \tan^2{(\pi \Phi/\Phi_0)}}$ are the charging and
Josephson energies of the transmon, respectively; here, $\Phi_0 = h / 2e $ is
the magnetic flux quantum, and critical current asymmetry in the Superconducting
QUantum Interference Device (SQUID) junctions is accounted for by the parameter
$d$.
The static dependence of the electron temperature $T_\mathrm{D}$ in the drain
(D) reservoir is determined by the temperature of the source (S) reservoir
$T_\mathrm{S}$ and the qubit detuning with respect to the resonators.
%

Figure~\ref{fig:system}b summarises the thermal model between the source and
drain reservoirs.  By voltage-biasing a pair of normal metal-insulator-superconductor
junctions (SINIS) attached to the source reservoir one can control its
temperature.
At sub-gap voltages, evacuation of hot quasiparticles from the source reservoir
lowers $T_\mathrm{S}$  below its unbiased value and above it, the biasing
provides conventional Joule heating~\cite{revmodgiaz}.  Under fixed experimental
conditions, the electrons in each normal-metal reservoir are in local thermal
equilibrium. 
In the detailed thermal balance we consider the interaction of the electron
system with the environment (resulting in an effective power $P_\mathrm{env}$
which includes the influence of SINIS biasing where appropriate) and with the
phonon bath (whose temperature $T_\mathrm{ph}$ is assumed to be uniform and
equal to the temperature of the cryostat). 
The latter mechanism is modeled by the conventional normal-metal electron-phonon
interaction $P_\mathrm{el-ph} = \Sigma \mathcal{V} (T_\mathrm{el}^5 -
T_\mathrm{ph}^5)$ that for small temperature differences can be linearised with
the thermal conductance $G_\mathrm{el-ph} = 5 \Sigma \mathcal{V}
T_\mathrm{el}^4$.  Here, $\mathcal{V}$ is the volume of the normal-metal
reservoir and $\Sigma$ is the corresponding electron-phonon coupling constant.
In the experiment, the source-to-drain heating power ($P_\mathrm{D} =
-P_\mathrm{S}$ by energy conservation) is determined by the response in $T_D$
under the assumption of the electron-phonon interaction dominating the thermal
relaxation of the electrons in the drain reservoir.
The lumped-element circuit representing the device is schematically
illustrated in Figure~\ref{fig:system}c.  
Each resonator is terminated at one end by a capacitor to
the transmon ($C_g \approx 8.6\, \mathrm{fF}$) and at the other end by the
normal-metal resistor to the Nb ground plane (Figure~\ref{fig:system}d).  This
configuration results in a quarter-wave resonator, with expected eigenfrequency
$f_\mathrm{r} = 6.4\,\mathrm{GHz}$ and quality factor $Q_\mathrm{r} = Z_0/R_N
\approx 20$, where $Z_0 =  \pi Z_{\infty}/4$ is the resonance impedance and
$Z_{\infty} = 50\, \Omega$ is the design impedance of the coplanar waveguide.
Here $R_N
\approx 2\, \mathrm{\Omega}$ is the nominal resistance of the N termination;
depending on the transparency of the metallic interfaces, additional dissipation
can significantly decrease the effective quality factor.
In our design the relaxation to the reservoir is the dominant source of losses
in the resonator, so that its quality factor $Q_\mathrm{r} \equiv 1/\gamma$. 

In modeling the system, one can consider the photonic reservoir-reservoir
coupling to be relatively weak, which allows us to apply standard perturbation
theory to describe it.  We expect the total thermal conductance between the
reservoirs to be three orders of magnitude lower than the quantum of thermal
conductance of a single channel $G_\mathrm{Q} = (\pi k_B^2/6\hbar) T$ at
temperature $T$~\cite{pendry_quantum_1983,schmidt_photon-mediated_2004}.  Here
we explore two photonic weak-coupling models, each based on the formalism
appropriate to the impact of reservoir-induced dissipation compared to the qubit
coupling rate.
We call these the quasi-Hamiltonian (QH) model for $\gamma \simeq g$ ,
and non-Hamiltonian (NH) model applicable when $\gamma \gg g$, respectively.
Conceptually, these two models showcase a different location for the Heisenberg
cut (\ie, the separation between the quantum subsystem and its classical
environment): either at the \emph{qubit-resonator to reservoir} boundaries
or at the \emph{qubit to resonator} interfaces, respectively.
In both models, the power to each reservoir is given by
\begin{equation}
	P_\mathrm{S/D} = \sum\limits_{k,l}\, \rho_{kk}\, E_{kl}\, \Gamma_{k \to
	l,\, \mathrm{S/D}} \, ,
	\label{eqn:powerbath}
\end{equation}
where $\rho$ is the density matrix and $E_{k,l}$, $\Gamma_{k \to l,\,
\mathrm{S/D}}$ are
the transition energy and rate for each respective reservoir, and the sum
runs over all the eigenstate indices $k,\,l$.  

\begin{figure}
	\includegraphics{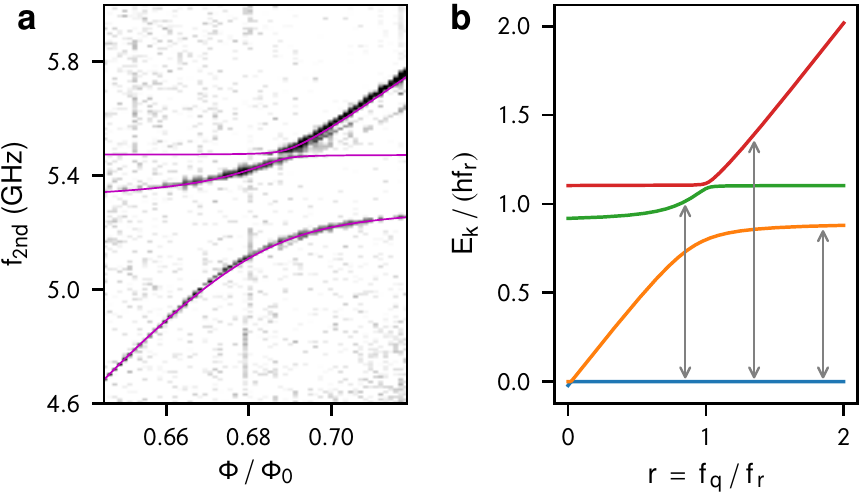}
	\caption{\textbf{Fundamental excitations of the
	resonator-qubit-resonator assembly.}
	\textbf{a}: Two-tone transmission spectroscopy data centered in
	the $f_\mathrm{r}
	\approx f_\mathrm{q}$ region for a sample in the fully-Hamiltonian limit. 
	Thin lines, representing
	eigenvalues derived from Equation~\ref{eqn:matrix}, are superimposed to
	the experimental dataset; optimal matching is obtained with parameters
	listed under column FH in supplementary information Table~1 with the addition
	of resonator asymmetry $a=0.008$ to reproduce the secondary avoided
	crossing visible at $f_\mathrm{2nd} \approx 5.47\,\mathrm{GHz}$.
	\textbf{b}: Overview of the dominant steady-state transitions between eigenstates 
	contributing to heat transport in the quasi-Hamiltonian limit according to
	Equation~\ref{eqn:MJCpower}. For visual clarity, here
	$g = -\tilde{g} = 0.1, \, a=0.05$.
	}
	\label{fig:spectroscopy}
\end{figure}

In the absence of dissipation, a \textbf{fully-Hamiltonian (FH)} description
 considers that the qubit and the two
resonators form a system of three coupled harmonic oscillators with level spacing
$h f_\mathrm{q}, \, h f_\mathrm{r}$. 
This neglects both nonlinear SQUID dynamics and occupation of higher resonator harmonics,
under the justification of quasi-static qubit drive and low temperatures in the two
reservoirs [$\beta_\mathrm{S/D} h f_\mathrm{r} \equiv h
f_\mathrm{r}/(k_BT_\mathrm{S/D}) \gg 1$], respectively.
The second-quantized Hamiltonian of the hybrid system reads
\begin{equation}
	\hat H = h f_\mathrm{r} \left[ ( \hat a_\mathrm{D}^\dagger \hat a_\mathrm{D} +
	\hat a_\mathrm{S}^\dagger \hat a_\mathrm{S})
	+ r \hat b^\dagger \hat b + 
	g ( \hat b \hat a_\mathrm{D}^\dagger + \hat b^\dagger \hat a_\mathrm{D} +
	\hat b \hat a_\mathrm{S}^\dagger + \hat b^\dagger \hat a_\mathrm{S} ) 
	+ \tilde{g} ( \hat a_\mathrm{D} \hat a_\mathrm{S}^\dagger + \hat
	a_\mathrm{D}^\dagger \hat a_\mathrm{S})
	\right] \, ,
\end{equation}
where $\tilde{g}$ quantifies direct resonator-to-resonator coupling.
Following the
low-temperature argument above, we choose the minimal four-level basis of 
$\{ \ket{000},\, \ket{100} ,\, \ket{010} ,\, \ket{001} \}$, where the entries in
each state refer to the S-resonator, the qubit, and the D-resonator,
respectively. 
With the addition of the parameter $a=\Delta f/f_\mathrm{r} \ll 1$ (quantifying
possible minor asymmetry $\Delta f$ between the eigenfrequencies of the two resonators),
this choice of basis results in the matrix representation
\begin{equation}
	\hat H = h f_\mathrm{r}
\begin{pmatrix}
	0 & 0 & 0 & 0 \\
	0 & 1+a/2 & g & \tilde{g} \\
	0 & g & r & g \\
	0 & \tilde{g} & g & 1-a/2
\end{pmatrix}\,.
	\label{eqn:matrix}
\end{equation}
In the $a\to 0$, $r\to 0$ limit, the photon cavity modes contribute a pair of eigenstates
corresponding to the symmetric and antisymmetric combinations of the eigenmodes
localized in each resonator. They are in general non-degenerate due to $\tilde{g}\neq 0$, 
and only the symmetric combination interacts with the qubit via $g$.
These features are evident in the dispersion of the eigenenergies shown in
Figure~\ref{fig:spectroscopy}b, where the dominant transitions between the
levels are also indicated.
To directly probe the flux-dependent spectrum of eigenstates
of the QHV in the FH limit ($\gamma \ll g$), we use a design
where the CPWs in the source and drain resonators are
connected directly to the ground plane without resistors. In this design, a
diagnostic resonator ($f_\mathrm{d} \approx 7.4\,\mathrm{GHz}$) is capacitively
coupled ($C_\mathrm{d} \approx 3.4\,\mathrm{fF}$) to the top arm of the transmon
island and inductively coupled to a microwave feedline.
Typical two-tone spectroscopic data, obtained by standard~\cite{bianchetti}
transmission readout of the diagnostic resonator is shown in
Figure~\ref{fig:spectroscopy}a. 
Inspection of the transition branches indicates
that the coupling capacitance $C_\mathrm{g}$ induces a $216\,\mathrm{MHz}$-wide avoided
crossing with the symmetric resonator eigenmode, consistent with $g =
0.02$. Additionally, a small ($<1\%$) asymmetry in resonator eigenfrequencies allows the
interaction between the qubit and the antisymmetric S/D resonator eigenmode,
visible as a minor avoided crossing at $f_\mathrm{2nd}\approx
5.47\,\mathrm{GHz}$.
These figures set the typical power scale of the qubit-mediated
heat transfer to $h f_\mathrm{r}^2 g \approx 0.4\,\mathrm{fW}$.

\begin{figure}
	\includegraphics{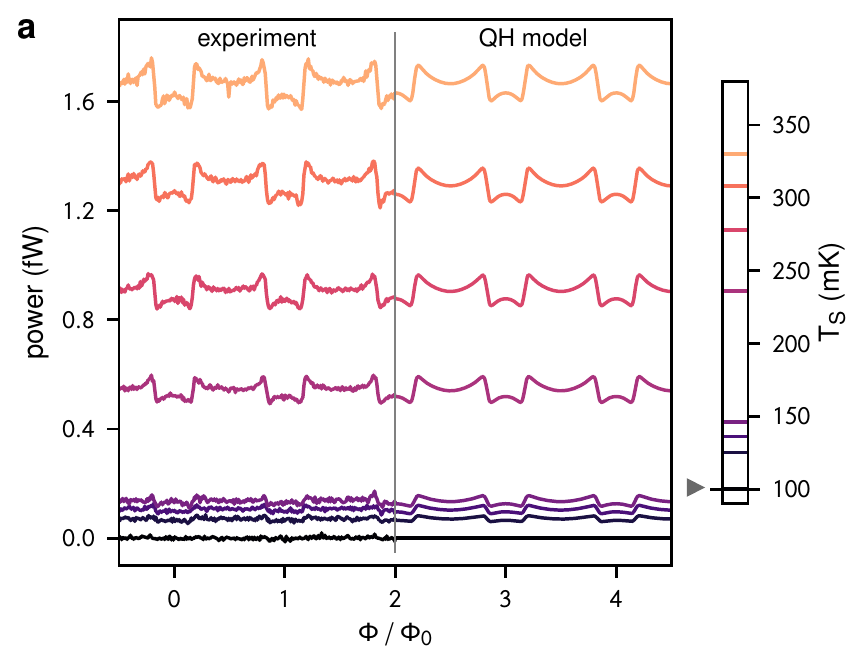}\\
	\includegraphics{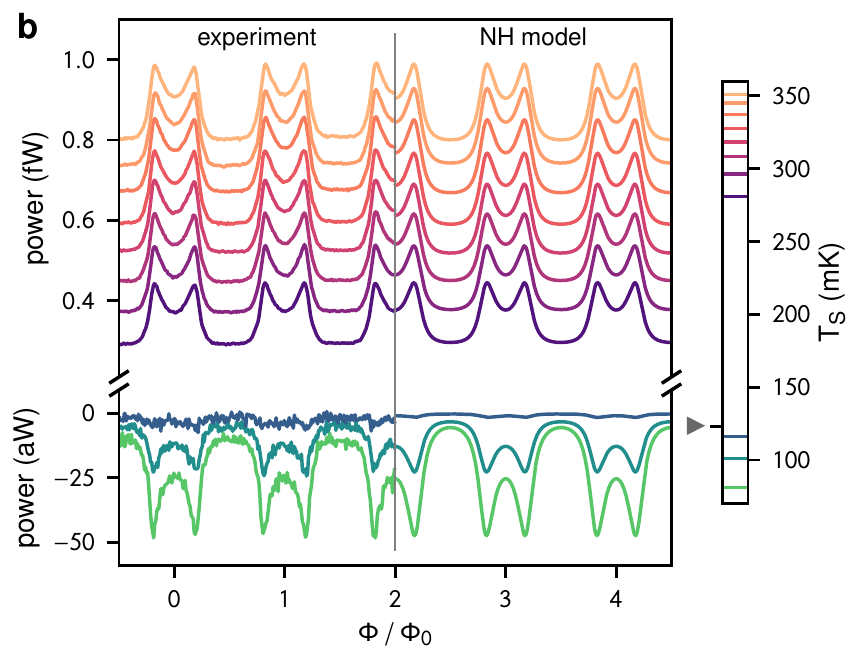}
	\caption{
	\textbf{Modulation of photonic heat transport.}
	Total heating power absorbed by the
	drain reservoir as a function of the applied magnetic flux $\Phi$.
	Different traces correspond to the source temperature values
	$T_\mathrm{S}$ shown in the adjacent legend bar. The unbiased
	temperature of the drain reservoir is here marked by a triangle.
	In each plot, experimental data is juxtaposed to the optimal fit
	of the appropriate theoretical model.
	Panels \textbf{a}, \textbf{b} correspond to
	quasi-Hamiltonian (Equation~\ref{eqn:MJCpower}) and
	non-Hamiltonian (Equation~\ref{eqn:spectralfilter}) regimes,
	respectively. Relevant modeling parameters are listed in supplemantary
	information Table~1.
        Residual reservoir-reservoir coupling, mediated by weak on-chip
	thermal conductance, is represented by an additional power-law
	contribution $P_0 =\xi [(T_\mathrm{S}/T_\mathrm{D})^n -1]$, where
	$\xi=5.14\,\mathrm{aW}$ and $n=4.63$ are empirical
	parameters.
	}
	\label{fig:experiment}
\end{figure}

We now consider the effect of introducing moderate dissipation to the system via
the S/D reservoirs, \ie, the \textbf{quasi-Hamiltonian (QH) regime}.
Equation~\ref{eqn:powerbath} allows us to determine the power from the
S-reservoir to the D-reservoir as
\begin{equation}
	P_\mathrm{D} = \frac{2 \pi h f_\mathrm{r}^2}{Q_\mathrm{r}} \sum_{k,\,l} \rho_{kk}
	\frac{| \bra{k} \hat a_\mathrm{D} - \hat a_\mathrm{D}^\dagger \ket{l} |^2}{
		1 + Q_\mathrm{r}^2 \left( f_{kl}/f_\mathrm{r} - f_\mathrm{r}/f_{kl} \right)^2 }\,
	\frac{(E_{kl}/h f_\mathrm{r})^2}{1-e^{-\beta_\mathrm{D} E_{kl}}} \,.
	\label{eqn:MJCpower}
\end{equation}
Here, the steady-state balance of the transition rates
$\Gamma_{k\to l}$ determines the level populations $\rho_{kk}$.
In this model, $P_{\rm D}$ is: 
i) limited by the reciprocal quality factor $Q_{\rm r}^{-1}\equiv \gamma$;
ii) non-vanishing at all values of flux even far away from the resonance;
iii) affected, around $f_\mathrm{q} =f_\mathrm{r}$, by
fast variation of populations, energy splitting, and matrix elements.
Experimental data for a QH-type sample recorded at $T_\mathrm{ph} = 45\,
\mathrm{mK}$ is presented in Figure~\ref{fig:experiment}a.
Here, different traces, representing the estimate for the 
power absorbed by the drain reservoir,
correspond to different thermal biases applied between the 
source reservoir ($T_\mathrm{S}$, controlled in the $100 \to 330\, \mathrm{mK}$
range) and the drain reservoir (unbiased temperature $T_\mathrm{D} \approx 100\, \mathrm{mK}$). 
The traces show a sizeable amount of
flux-independent power transmitted to the drain reservoir. This is particularly
impressive for complete resonator-qubit detuning for applied flux corresponding
to half-integer values of $\Phi_0$. The origin of this power flow between the
reservoirs lies in the role of the two mixed S/D resonator eigenmodes
spanning the whole resonator-qubit-resonator assembly.
Remarkably, approaching the $f_\mathrm{q} > f_\mathrm{r}$ condition near integer flux bias values
results in an initial increase of the absorbed power, followed by a step-like
decrease and a partial revival when reaching integer $\Phi / \Phi_0$
values, where $f_\mathrm{q}(\Phi)$ is maximal.
The comparison with the theoretical prediction provided by
Equation~\ref{eqn:MJCpower} with the nominal $Q_\mathrm{r}=20$ value indicates that the
model captures all these features quantitatively. In this case, optimal
reproduction of experimental data is found, according to the estimates presented in
supplementary information Table~1, with $g \approx 0.019$ and $\tilde{g} \approx -0.020$.
These values compare well to the ones directly measured from the two-tone
spectroscopy of the FH-type samples; notably, 
$g/\gamma = gQ_\mathrm{r} \approx 0.4$.

\textbf{The NH model} is described in Ref.~\cite{karimi_otto_2016}.
The power from S-reservoir to the D-reservoir reads
\begin{equation}\label{eqn:spectralfilter}
\begin{split}
P_\mathrm{D} =\, &\pi h g f_\mathrm{r}^2 \frac{n(\beta_\mathrm{S} h f_\mathrm{q}) -
	n(\beta_\mathrm{D} h f_\mathrm{q})}{[1 + Q_\mathrm{r}^2 ( r - 1/r)^2] [
		\coth( {\beta_\mathrm{S} h f_\mathrm{q}}/{2}) +\coth (
		{\beta_\mathrm{D} h f_\mathrm{q}}/{2} ) ]} \\
	+\, & \pi h \kappa f_\mathrm{r}^2 \int_0^\infty 
	\frac{n(x\,\beta_\mathrm{S} h f_\mathrm{r}) - n(x\,\beta_\mathrm{D} h
	f_\mathrm{r})}{[1 + Q_\mathrm{r}^2 (
	x - 1/x)^2]^2}
	x^3 \mathrm{d}x
\end{split} \, ,
\end{equation}
where $n(\beta_\mathrm{S/D} h f) = 1/(\exp(\beta_\mathrm{S/D} h f) -1 )$ is the
equilibrium mode population in each resonator; the
second term describes direct resonator-to-resonator photon transfer, quantified
by $\kappa$.  Overall, $P_\mathrm{D}$ is:
i) limited by the couplings $g,\,\kappa$ (as opposed to $\gamma$ in the QH
model);
ii) peaking when the qubit transition frequency matches the resonator eigenfrequency,
$f_\mathrm{q}=f_\mathrm{r}$;
iii) inhibited when the qubit-resonator detuning exceeds the resonator linewidth
$|f_\mathrm{q}-f_\mathrm{r}|/f_\mathrm{r} \gg Q_\mathrm{r}^{-1}$. 
The flux dependence of the power to the D-reservoir recorded at $T_\mathrm{ph}=55\, \mathrm{mK}$
for an NH-design device is shown in Figure~\ref{fig:experiment}b.
This dependence is consistent with the expectations based on
Equation~\ref{eqn:spectralfilter}.
Here, different traces, representing the estimate of the 
power absorbed by the drain reservoir,
correspond to different thermal biases applied between the 
source reservoir ($T_\mathrm{S}$, controlled in the $80 \to 360\, \mathrm{mK}$
range) and the drain reservoir (unbiased temperature $T_\mathrm{D} \approx 120\, \mathrm{mK}$). 
The modulation of all traces shows clear presence of two broad peaks per
flux period, corresponding to the condition $f_\mathrm{q} = f_\mathrm{r}$.
The shape of the flux modulation appears independent of the sign of the thermal
bias. This sign reversal can be observed in the traces corresponding to the three lowest
values of $T_\mathrm{S}$, obtained by electron-cooling (instead of heating) of
the source reservoir with an appropriate sub-gap voltage bias of a SINIS junction pair.
In this figure, experimental data
is juxtaposed to the best fit of Equation~\ref{eqn:spectralfilter} yielding
the parameter estimates listed in supplementary information Table~1,
in particular $Q_\mathrm{r} = 3.15 \pm 0.14$. 
Such a low quality factor fully justifies the adoption of the
NH model, even in the presence of a non-negligible coupling: $g/\gamma =
gQ_\mathrm{r} \approx 0.05$.

\begin{figure}
	\includegraphics{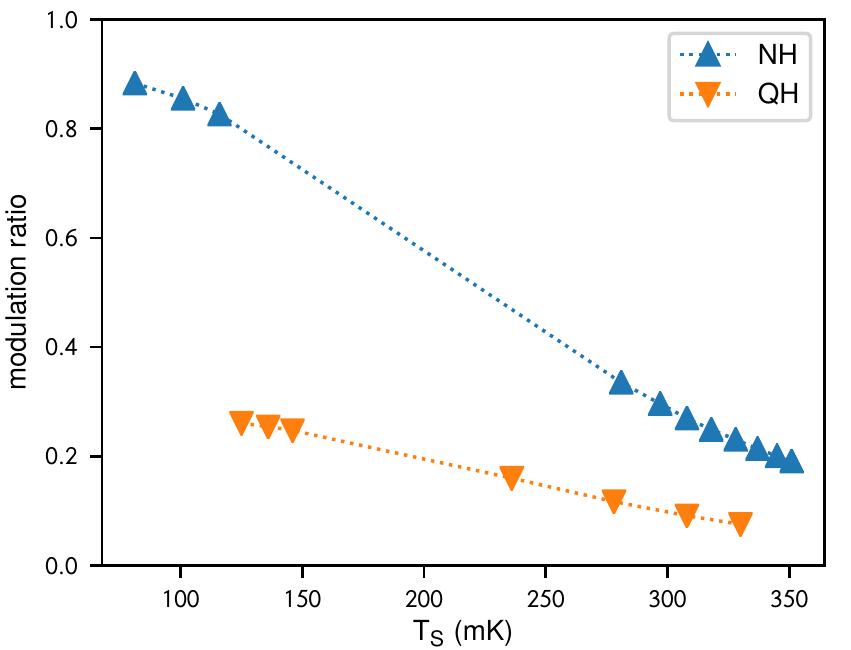}
	\caption{
	\textbf{Quantum Heat Valve performance.}
	Valve modulation ratio $(\max_\Phi P_\mathrm{D} - \min_\Phi
	P_\mathrm{D}) / \max_\Phi |P_\mathrm{D}|$ as a function of the source
	temperature $T_\mathrm{S}$ for non-Hamiltonian and quasi-Hamiltonian
	regimes. Dotted lines are intended as a visual aid.}
	\label{fig:ratio}
\end{figure}

In the NH case, the number of photonic excitations in each resonator is
dominated by dissipative processes in the reservoirs.
Under this hypothesis, the overdamping prevents the formation of the mixed
S/D eigenmodes characteristic of the Hamiltonian limit. Notably, the presence of these
excitations (spectroscopically probed in the FH sample, where $g/\gamma \gg 1$),
is required to quantitatively reproduce via Equation~\ref{eqn:MJCpower}
the heat modulation observed in the QH sample,
in spite of its arguably low $g/\gamma \approx 0.4$.
In the NH limit, instead, the excitation of the qubit acts as an independent
flux-tunable spectral filter between the photonic populations tied to the
source and drain reservoirs (Equation~\ref{eqn:spectralfilter}).   
Figure~\ref{fig:ratio} presents a comparison of QH and NH samples in terms
of performance as a heat valve.
We see that the highest modulation ratio is obtained in the NH sample for low
temperatures, where the flux-independent ``background'' contributions are small in
comparison to the actual photonic power.

The Quantum Heat Valve presented here is a key platform
dedicated to the investigation of quantum thermodynamic phenomena in
hybrid mesoscopic/circuit QED systems.
Planning devices including active thermal degrees of
freedom requires matching resonator eigenenergies to the expected reservoir
temperature. The principal heat transport bottleneck can be the resonator-qubit
coupling, typically $g \lesssim 0.05$ for coplanar elements.  On the other hand,
a comparably strong resonator-reservoir relaxation mechanism is required for the
thermalization of the relevant photonic mode population.  We find that the
competition between qubit-resonator and reservoir-resonator couplings affects
strongly not only the power scale of the heat transport, but also the locality of 
its physical origin.

\subsection*{Methods}
\emph{Fabrication protocols}
The devices were fabricated on $330\,\micron$ thick sapphire substrates coated with
$200\,\mathrm{nm}$-thick sputtered niobium film. Broader
features, such as coplanar waveguides, transmon island and electrode
fanout were patterned by reactive ion etching on an electron-beam
lithography-defined mask. 
The CPW design
features a $20\,\mathrm{\mu m}$-wide centreline spaced by $10\, \mathrm{\mu m}$
with respect to the ground plane, resulting in capacitance and inductance per
unit length of $153\,\mathrm{fF/mm}$ and $403\,\mathrm{pH/mm}$ respectively.
All chip layouts are available in the Supplementary Information.
Nanostructures including the tunnel junction elements were
realised in two steps with shadow-mask electron-beam lithography on a
$1\,\micron$-thick poly(methyl-metacrylate) / copolymer resist bilayer, followed
by tilted thin film deposition in an electron-beam evaporator.
In the first step, two offset depositions of $28\,\mathrm{nm}$-thick Al layers
(with intermediate oxidation) are performed to realise the transmon SQUID
 (Figure \ref{fig:system}e) with typical per-junction tunnel
resistance $R_\mathrm{T} \approx 7\, \mathrm{k\Omega}$ at cryogenic
temperature. 
In the second step, the terminations of the resonators are realised by first
depositing and oxidising a $15\,\mathrm{nm}$-thick Al layer, followed by a
$50\,\mathrm{nm}$ Cu layer and finally by a $85\,\mathrm{nm}$-thick Al layer in
clean contact with the Cu layer.
The typical tunnel resistance is $R_\mathrm{NIS} \approx
25\,\mathrm{k\Omega}$; during the experiment, these electrodes are connected to fanout lines
for the setting and readout of the electron temperature in the reservoirs over a
typical timescale of tens of milliseconds.
In both steps, the contact between the Nb substrate and the deposited
metal is facilitated by \textit{in situ} Ar ion plasma milling, while
tunnel junctions are realised by controlled
oxidation (oxygen partial pressure $\approx 10\,\mathrm{mbar}$ for $8\,\mathrm{min}$).
After liftoff in acetone and cleaning in isopropyl alcohol, the substrates are
diced to size ($4 \times 8 \, \mathrm{mm}$ for QH/NH-type
and $7 \times 7 \, \mathrm{mm}$ for FH-type chips) with a
diamond-coated resin blade and wire-bonded to a custom made brass chip carrier
for the cryogenic characterisation.

\emph{Measurements}
The experiment has been performed in a custom-made dilution
refrigerator able to reach base temperature values $\approx
50\, \mathrm{mK}$. The bonded chip, shielded by two brass Faraday enclosures, is
connected to the room-temperature breakout box with conventional cryogenic
signal lines. Each line is filtered by a
$1\,\mathrm{m}$-long Thermocoax wire segment,
resulting in an effective signal bandwidth of $0-10\, \mathrm{kHz}$, for
low-impedance loads.
Magnetic field is applied perpendicular to the sample substrate by a
superconducting magnet wound on the exterior of the insert vacuum can. The
latter is inserted in a high-permeability magnetic shield.

Current and voltage electrical bias are applied by programmable voltage sources
and function generators with appropriate room-temperature resistor networks. 
Current and voltage amplification is performed by room-temperature low-noise
amplifiers (FEMTO Messtechnik GmbH, models DLPCA-200 and DLPVA-100).
In order to minimise the impact of signal pickup and low-frequency drifts of the
differential voltage amplifier output in the SINIS thermometer readout, 
temperature signals are derived from the first harmonic recorded by a lock-in
amplifier synchronised to the square-wave modulation ($42\, \mathrm{Hz}$)
of the voltage bias of the source reservoir SINIS circuit.
The noise-equivalent spectral density obtained in this
differential readout scheme is $0.1\, \mathrm{mK/\sqrt{Hz}}$, corresponding to typical 
uncertainty $\delta T \approx 40\, \mu\mathrm{K}$~(r.m.s.) in the temperature estimates
(effective integration bandwidth $ = 0.14\, \mathrm{Hz}$ for the lock-in measurement).
Quantitative estimates of the bias-dependent power absorbed by the drain
reservoir are obtained assuming that the electron-phonon interaction dominates
the thermal relaxation, yielding
\begin{equation}
	P_\mathrm{D} \approx \Sigma \mathcal{V} T_\mathrm{D}^4 \Delta T_\mathrm{D} \, ,
\end{equation}
where $\Delta T_\mathrm{D}$ is the peak-peak amplitude of the signal recorded by the drain
thermometer.

\section*{Acknowledgements}
This work was funded through Academy of Finland grants 297240, 312057
and 303677 and from the European Union's Horizon 2020
research and innovation programme under the European Research Council (ERC)
programme and Marie Sklodowska-Curie actions (grant agreements 742559 and
766025). 
This work was supported by Centre for Quantum Engineering (CQE) at Aalto
University.
We acknowledge the facilities and technical support of Otaniemi research
infrastructure for Micro and Nanotechnologies  (OtaNano), and VTT technical
research center for sputtered Nb films.
We acknowledge M. Meschke for technical help and O-P. Saira for useful
discussions in the initial stages of this work. We thank D. Golubev and
Y. Galperin for helpful discussions.

\section*{Author contributions}
The experiment was conceived by J.P. and B.K., with contributions from CD.C.
A.R. performed the experiment. A.R., J.S. and Y-C.C. designed and fabricated the
samples. Data analysis was performed by A.R. based on theoretical models
conceived and solved by J.P. and B.K. Y.-C.C. performed the spectroscopy
measurements. J.T.P. provided technical support in fabrication,
low-temperature setups and measurements. All authors have been involved in the
discussion of scientific results and implications of this work. The manuscript
was written by A.R. with contributions from J.P., B.K., and J.S.

\newpage
\section*{Supplementary information}
\subsection*{Chip design} \label{section1}
\begin{figure}[th]
\centering
\includegraphics [width=0.5\columnwidth] {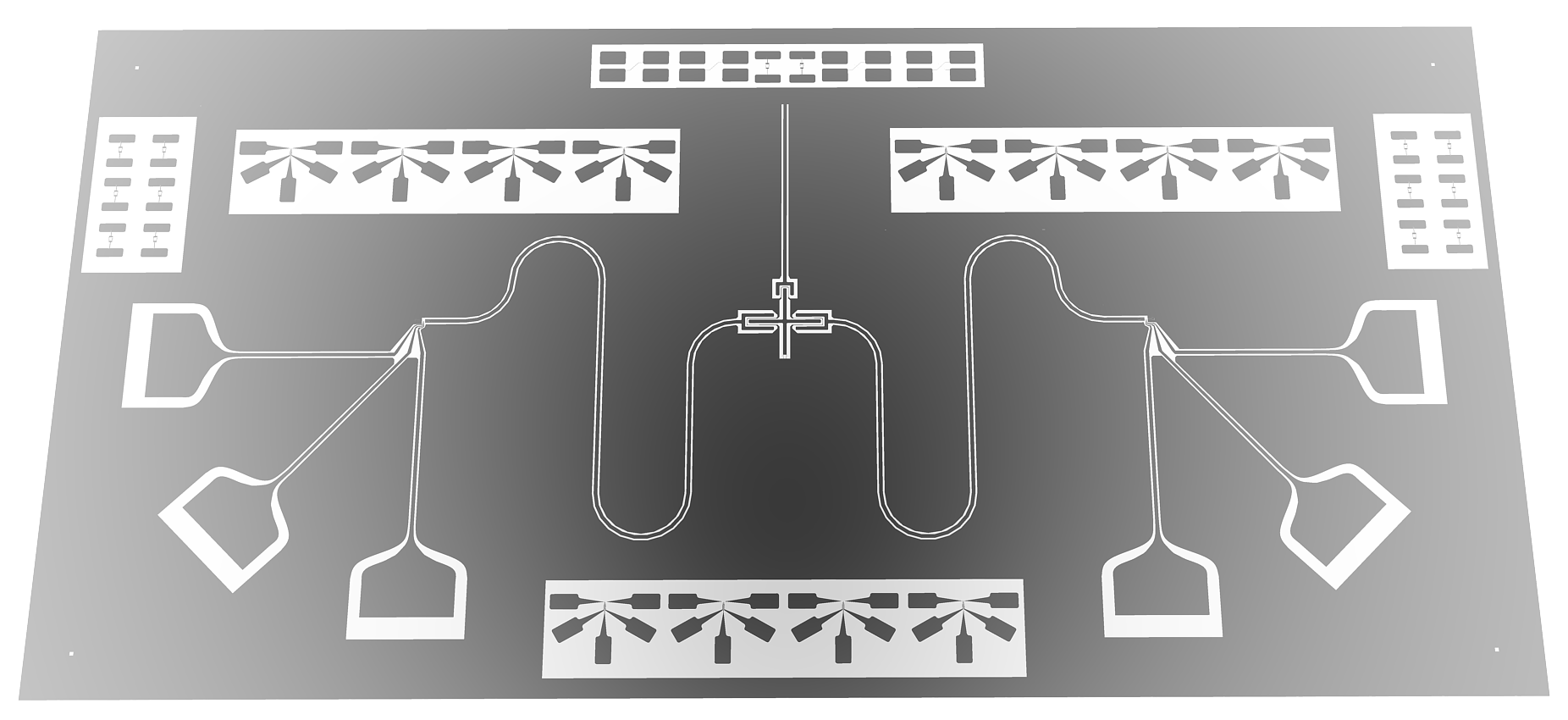}\\
\includegraphics [width=0.5\columnwidth] {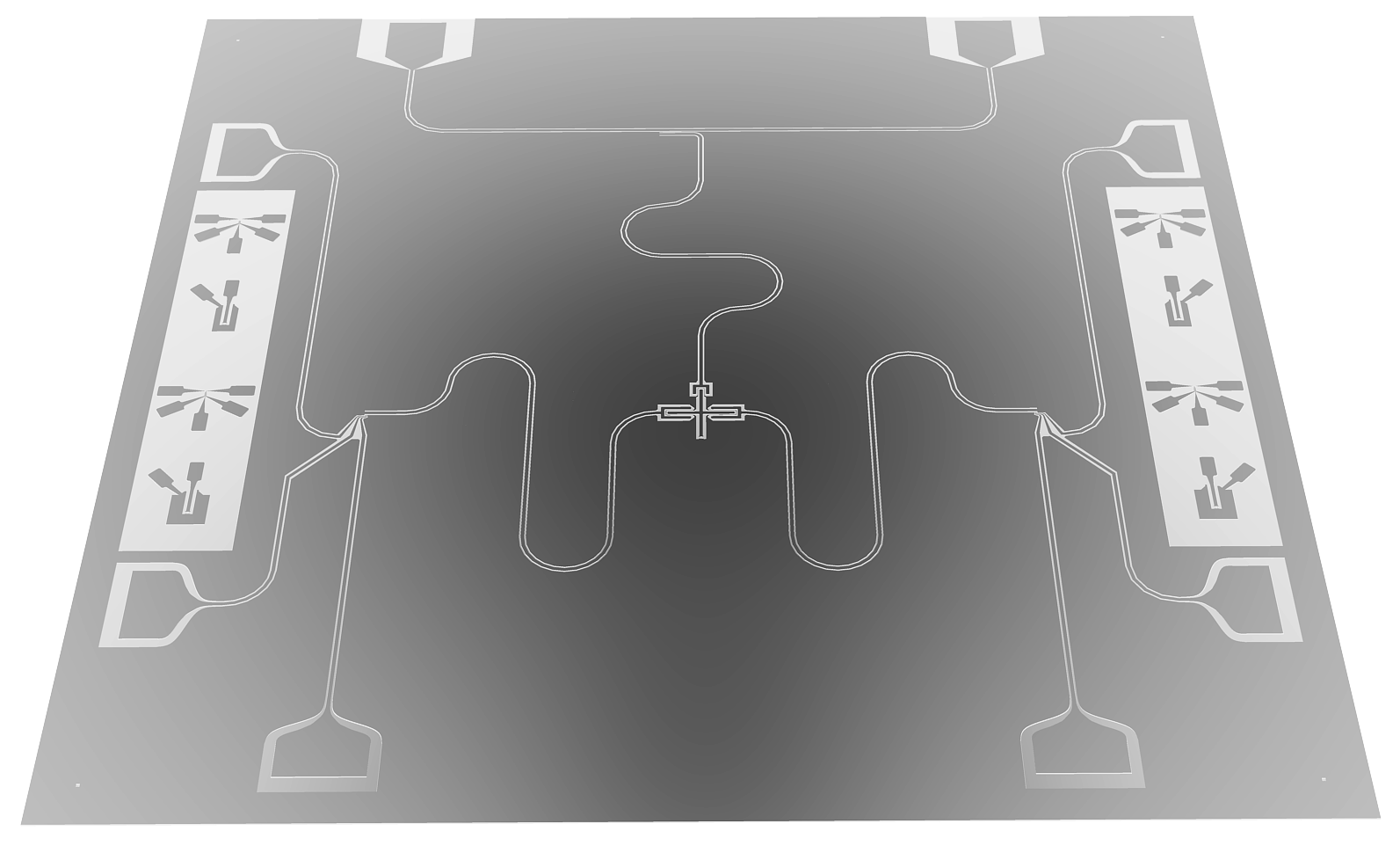}
\caption{Renders of the coplanar microwave structure of a quantum heat valve
	device.  The dark areas represent metallic Nb film after the etching
	step.  The additional structures at the periphery are fabricated for
	diagnostic purposes and are not part of the actual device. 
	 The \textbf{top image} refers to samples of type QH/NH in the main text
	 (4x8 mm). Here the cross-shaped transmon element is coupled to
	 identical quarter-wave resonators (left and right), each of which is
	 terminated by a normal-metal Cu shunt to the common ground plane.  The
	 three contact pads near the bottom corners of the device are connected
	 to NIS probes on the normal-metal terminations of the resonators.
	 The \textbf{bottom image} refers instead to FH-type samples (7x7 mm). In
	 this design, each quarter-wave resonator is directly connected to the
	 ground plane, and a $7.4\,\mathrm{GHz}$ diagnostic resonator couples
	 the top terminal of the transmon structure to a feedline for
	 spectroscopy characterization via transmission microwave readout. In
	 this design, the fanout for the probe electrodes is still patterned in
	 the Nb film, but no actual NIS elements are present near the
	 quarter-wave shunt terminations.
\label{fig1}}
\end{figure}

A rendered image of the quantum heat valve is shown in Fig.~\ref{fig1}. 

\subsection*{Device parameters}
A summary of device parameters is presented in Tab.~\ref{tab:params}.
\begin{table}
\begin{tabular}{l|cccc}
\toprule
	& \textbf{design} & \textbf{FH} & \textbf{QH} & \textbf{NH} \\
\midrule
$f_\mathrm{r}$ (GHz)    & $6.4$ & $5.39$ & $5.30 \pm 0.04$ & $5.61 \pm 0.15$ \\
$Q_\mathrm{r}$          & $20$  & N/A    & $20$            & $3.15 \pm 0.14$ \\
$g\times 10^2$          & $2.0$ & $2.0$  & $1.93 \pm 0.02$ & $1.56 \pm 0.06$ \\
	$\tilde{g}\times 10^2$  & N/A   & $-1.5$ & $-2.01\pm 0.05$ & N/A \\
	$\kappa\times 10^2$  & N/A   & N/A & N/A & $0.21 \pm 0.05$ \\
$E_\mathrm{C}/h$ (GHz)  & $0.2$ & $0.15$ & $0.15$          & $0.15$ \\
$E_\mathrm{J0}/h$ (GHz) & $45$  & $45.0$ & $28.8 \pm 0.3$  & $35.7 \pm 1.1$ \\
$d$                     & N/A   & N/A    & $0.57 \pm 0.01$ & $0.08 \pm 0.59$ \\
\bottomrule
\end{tabular}
	\caption{Summary of parameter estimates and measurements.
Values quoted with uncertainty are parameter estimates obtained by fitting
Equations~5 and~6 (main article text) to experimental data
recorded for samples QH and NH, respectively. Values quoted without uncertainty,
either by design or measured on FH-type samples, have been used as 
constraints in modeling samples QH, NH. }
	\label{tab:params}
\end{table}

\newpage
\subsection*{The Hamiltonian}

In the general case, we assume the qubit to be fully harmonic and
capacitively coupled ($g$) to two resonators, also modelled as harmonic oscillators. 
The two resonators have eigen-frequency and quality factor
$f_\mathrm{r},\,Q_\mathrm{r}$, respectively, while the energy splitting
of the harmonic potential of the transmon is $h f_\mathrm{q}$.
Since the non-Hamiltonian model for $gQ_\mathrm{r} \ll 1$ is described elsewhere \cite{Jukka1,Jukka2},
we focus here only on the quasi-Hamiltonian applicable when $gQ_\mathrm{r} \gtrsim 1$. 
\begin{figure}[th]
\centering
\includegraphics [width=0.5\columnwidth] {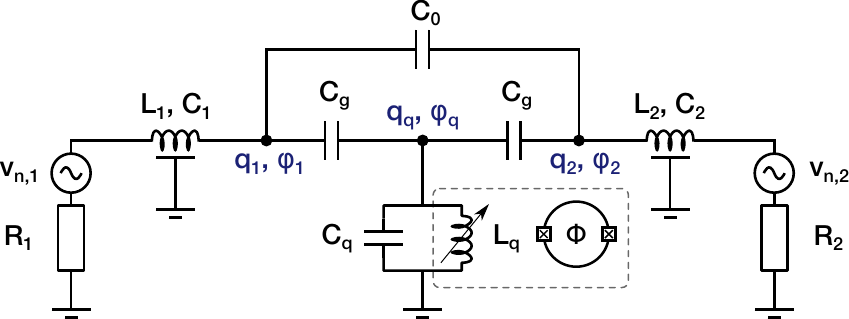}
\caption{Lumped-element approximation of the quantum heat valve. The notations
	for phases $\varphi_{\rm i}$, charges $q_{\rm i}$, capacitances $C_{\rm
	i}$, inductances $L_{\rm i}$, and resistances $R_{\rm i}$ in the text
	can be read in this figure.
\label{fig2}}
\end{figure}

Figure \ref{fig2} shows a lumped-element approximation of the system.
In particular, to include an example of a direct resonator-resonator photon
transfer mechanism, we introduce the shunting capacitor $C_0$.
The Lagrangian of the
circuit consisting of parallel $LC$ resonators in this case reads then
\begin{equation}\label{totallagrangian}
\mathfrak{L}(\varphi_{\rm 1}, \dot{\varphi}_{\rm 1}, \varphi_{\rm q}, \dot{\varphi}_{\rm q}, \varphi_{\rm 2}, \dot{\varphi}_{\rm 2})=\frac{1}{2}\big{(}C_{\rm 1}\dot{\varphi}_{\rm 1}^2+C_{\rm g}(\dot{\varphi}_{\rm q}-\dot{\varphi}_{\rm 1})^2+C_{\rm q}\dot{\varphi}_{\rm q}^2+C_{\rm g}(\dot{\varphi}_{\rm q}-\dot{\varphi}_{\rm 2})^2+C_0 (\dot{\varphi}_{\rm 1}-\dot{\varphi}_{\rm 2})^2+C_{\rm 2}\dot{\varphi}_2^2\big{)}-\frac{1}{2}\big{(}\frac{\varphi_1^2}{L_1}+\frac{\varphi_{\rm q}^2}{L_{\rm q}}+\frac{\varphi_2^2}{L_2}\big{)}.
\end{equation}
From the Lagrangian we can obtain the conjugate momenta of node fluxes by the Legendre transformation $q_{\rm n}=\frac{\partial \mathfrak{L}}{\partial \dot{\varphi}_{\rm n}}$,
yielding:
\begin{equation}
H=(\frac{q_1^2}{2C_{\rm 1,eff}}+\frac{\varphi_1^2}{2L_1})+(\frac{q_{\rm q}^2}{2C_{\rm q,eff}}+\frac{\varphi_{\rm q}^2}{2L_{\rm q}})+(\frac{q_2^2}{2C_{\rm 2,eff}}+\frac{\varphi_2^2}{2L_2})+C_{\rm 1q}^{-1}q_{\rm q}q_1+C_{\rm 2q}^{-1}q_{\rm q}q_2+C_{12}^{-1}q_1q_2\, .
\end{equation}
Applying symmetry considerations for this system, namely $C_1=C_2\equiv C_{\rm r}$ and $L_1=L_2\equiv L_{\rm r}$, and for  $C_{\rm g}\ll C_{\rm r},C_{\rm q}$ (considering up to linear in $C_{\rm g}$ terms only)
\begin{eqnarray}
&&C_{\rm 1,eff}^{-1}=C_{\rm 2,eff}^{-1}=\frac{C_0+C_{\rm r}}{2C_0C_{\rm r}+C_{\rm r}^2}-C_{\rm g}\frac{2C_0^2+2C_0C_{\rm r}+C_{\rm r}^2}{(2C_0C_{\rm r}+C_{\rm r}^2)^2}\nonumber\\&&
C_{\rm q,eff}^{-1}=\frac{1}{C_{\rm q}}(1-\frac{2C_{\rm g}}{C_{\rm q}})\nonumber\\
&&C_{12}^{-1}=\frac{C_0}{2C_0C_{\rm r}+C_{\rm r}^2}-\frac{2C_0(C_0+C_{\rm r})C_g}{C_{\rm r}^2(2C_0+C_{\rm r})^2}\nonumber\\
&&C_{\rm c}^{-1} \equiv C_{\rm 1q}^{-1}=C_{\rm 2q}^{-1}=\frac{C_{\rm g}}{C_{\rm q}C_{\rm r}}.
\end{eqnarray}
The Hamiltonian based on the charge operators $\hat{q}_{\rm i}=-i\sqrt{\frac{\hbar}{2Z_0}}(\hat{a}_{\rm i}-\hat{a}_{\rm i}^\dagger)$ and $\hat{q}_{\rm q}=-i\sqrt{\frac{\hbar}{2Z_0}}(\hat{b}-\hat{b}^\dagger)$ is given by
\begin{equation}
(hf_{\rm r})^{-1}\hat{H}=\hat{a}_1^\dagger \hat{a}_1+\hat{a}_2^\dagger \hat{a}_2+r\hat{b}^\dagger \hat{b}+g(\hat{b}\hat{a}_1^\dagger+\hat{b}^\dagger\hat{a}_1+\hat{b}\hat{a}_2^\dagger+\hat{b}^\dagger\hat{a}_2)+\tilde{g}(\hat{a}_1\hat{a}_2^\dagger+\hat{a}_1^\dagger\hat{a}_2),
\end{equation}
where $g=(4\pi Z_0C_{\rm c}f_{\rm r})^{-1}$, $\tilde{g}=(4\pi Z_0C_{12}f_{\rm r})^{-1}$, $r=f_{\rm q}/f_{\rm r}$, and $Z_0=\sqrt{L_{\rm r}/C_{\rm r}}$. Here $\hat{a_{\rm i}}^\dagger,\hat{b}^\dagger$ and $\hat{a_{\rm i}},\hat{b}$ are the creation and annihilation operators, respectively. In the product basis $\{|000\rangle,$ $|100\rangle,$ $|010\rangle,$ $|001\rangle \}$ where the first entry refers to the left resonator, second one to the qubit, and the last one to the right resonator we have then the result
\begin{eqnarray} \label{hamiltonian3}
H = hf_r \left(
\begin{array}{cccc}
0 & 0 &0 & 0 \\ 0 & 1&  g & \tilde{g} \\ 0 & g & r & g \\  0 & \tilde{g} & g & 1
\end{array}
\right),
\end{eqnarray}

The dimensionless eigenenergies $\lambda_{k}=E_{k}/(hf_{\rm r})$ of this Hamiltonian are 
\begin{eqnarray}
&&\lambda_1=0\nonumber\\&&
\lambda_2=1-\tilde{g}\nonumber\\&&
\lambda_3=\frac{1}{2} [1+\tilde{g}+r-\sqrt{1+2\tilde{g}+\tilde{g}^2+8g^2-2r-2\tilde{g}r+r^2}]\nonumber\\&&
\lambda_4=\frac{1}{2} [1+\tilde{g}+r+\sqrt{1+2\tilde{g}+\tilde{g}^2+8g^2-2r-2\tilde{g}r+r^2}]
\end{eqnarray}
These energy levels, and the allowed transition rates between them are shown in Fig. 2b of the main text.

\subsection*{Coupling to thermal noise: transition rates and power} 
Consider first the resonator as a series $LC$ circuit, whose impedance is
\begin{equation} \label{h10}
Z=i Z_0(f/f_{\rm r}-f_{\rm r}/f).
\end{equation}
With series resistance $R$, the quality factor of the resonance is $Q_\mathrm{r}=Z_0/R$. 

In reality, the resonator is a $\lambda /4$ transmission line terminated by an open circuit (by a small gate capacitance). Its impedance is given by
\begin{equation} \label{h11}
Z=\frac{e^{2ikl}-\chi}{e^{2ikl}+\chi}Z_\infty,
\end{equation}  
where the reflection coefficient $\chi=(Z_\infty-Z_{\rm L})/(Z_\infty + Z_{\rm L}) \rightarrow -1$, when the load impedance $Z_{\rm L}\rightarrow \infty$. Here $Z_\infty$ equals $\sqrt{L_0/C_0}$ with $L_0,C_0$ the inductance and capacitance per unit length of the line. Near the resonance, i.e. for $k=k_0+\delta k$ values near $k_0l=\pi/2$, we may expand $Z$ with the result
\begin{equation} \label{h12}
Z\approx i\frac{\pi}{4}Z_\infty (f/f_{\rm r}-f_{\rm r}/f),
\end{equation}
where $k_0 = \omega_0 \sqrt{L_0C_0}$.
Then with $Q_\mathrm{r}=Z_0 /R$, where $Z_0=\frac{\pi}{4}Z_\infty$, we obtain identical results with the lumped series resonator above. In both cases, the fluctuating voltage seen by the harmonic oscillator is normalized to 
\begin{equation} \label{h10}
v(t) = \frac{R}{R+iZ_0(f/f_{\rm r}-f_{\rm r}/f)} v_{\rm n}(t),
\end{equation}  
where $v_{\rm n}(t)$ is the noise of the resistor alone with spectrum
\begin{equation} \label{h8}
S_{v_n}(f)= \frac{2Rhf}{1-e^{-\beta hf}}.
\end{equation} 
Then
\begin{equation} \label{h11}
	S_v(f)= \frac{1}{1+Q_\mathrm{r}^2 (f/f_{\rm r}-f_{\rm r}/f)^2}\frac{2Rhf}{1-e^{-\beta hf}}.
\end{equation}
We obtain the transition rates induced by thermal bath $\rm B$ between eigenstates $k$ and $l$ by the golden rule expression
\begin{equation} \label{h12}
\Gamma_{k\rightarrow l,{\rm B}} =\frac{1}{\hbar^2}|\langle k|\hat q_{\rm B}|l\rangle |^2 S_{v,\rm B}(f_{kl}).
\end{equation} 
Here $S_{v,\rm B}$ is the noise on the resonator coupled to bath $\rm B$. 

If the energy separations between the eigenenergies are given by $E_{kl}=hf_{\rm r}(\lambda_k-\lambda_l)\equiv hf_{kl}$, we then have
\begin{eqnarray} \label{h12b}
&&\Gamma_{k\rightarrow l,{\rm B}} = \frac{2\pi}{Q_{\rm B}}\frac{|\langle k|\hat{a}_{\rm B}-\hat{a}_{\rm B}^\dagger|l\rangle|^2}{1+Q_{\rm B}^2(\frac{f_{kl}}{f_{{\rm r}}}-\frac{f_{\rm r}}{f_{kl}})^2}\frac{f_{kl}}{1-e^{-\beta_{\rm B}hf_{kl}}},
\end{eqnarray}
where $Q_{\rm B}$ is the quality factor of the resonator attached to bath $\rm B$ at inverse temperature $\beta_{\rm B}$. Here the squared matrix elements are given by
\begin{eqnarray} \label{h14}
&&|\langle 1|\hat{a}_{\rm B}-\hat{a}_{\rm B}^\dagger|2\rangle|^2 =|\langle 2|\hat{a}_{\rm B}-\hat{a}_{\rm B}^\dagger|1\rangle|^2 =1/2 \nonumber \\&&
|\langle 1|\hat{a}_{\rm B}-\hat{a}_{\rm B}^\dagger|3\rangle|^2 =|\langle 3|\hat{a}_{\rm B}-\hat{a}_{\rm B}^\dagger|1\rangle|^2 =\frac{1}{4}(1+\frac{r-1-\tilde{g}}{\sqrt{(r-1-\tilde{g})^2+8 g^2}})
\nonumber \\&&
|\langle 1|\hat{a}_{\rm B}-\hat{a}_{\rm B}^\dagger|4\rangle|^2 =|\langle 4|\hat{a}_{\rm B}-\hat{a}_{\rm B}^\dagger|1\rangle|^2 =1/2- |\langle 1|\hat{a}_{\rm B}-\hat{a}_{\rm B}^\dagger|3\rangle|^2 = \frac{1}{4}(1-\frac{r-1-\tilde{g}}{\sqrt{(r-1-\tilde{g})^2+8 g^2}}).
\end{eqnarray}
Other elements vanish.

Based on the allowed transitions presented in Fig.~2b, the diagonal elements of the density matrix under non-driven conditions are given by
\begin{eqnarray}
&&\dot{\rho}_{11}=-(\Gamma_{1\rightarrow 2}+\Gamma_{1\rightarrow 3}+\Gamma_{1\rightarrow 4})\rho_{11}+\Gamma_{2\rightarrow 1}\rho_{22}+\Gamma_{3\rightarrow 1}\rho_{33}+\Gamma_{4\rightarrow 1}\rho_{44}\nonumber\\
&&\dot{\rho}_{22}=-\Gamma_{2\rightarrow 1}\rho_{22}+\Gamma_{1\rightarrow 2}\rho_{11}\nonumber\\
&&\dot{\rho}_{33}=-\Gamma_{3\rightarrow 1}\rho_{33}+\Gamma_{1\rightarrow 3}\rho_{11}\nonumber\\
&&\dot{\rho}_{44}=-\Gamma_{4\rightarrow 1}\rho_{44}+\Gamma_{1\rightarrow 4}\rho_{11},
\end{eqnarray}
where $\Gamma_{k\rightarrow l}=\Gamma_{k\rightarrow l,{\rm S}}+\Gamma_{k\rightarrow l,{\rm D}}$. We apply steady state conditions as $\dot{\rho}=0$. The power to bath $\rm B$ then reads
\begin{equation}\label{power}
	P_{\rm B}=\frac{2\pi h f_{\rm r}^2}{Q_{\rm B}}\sum_{k,l}\,\rho_{kk} \, \frac{|\langle
	k|\hat{a}_{\rm B}-\hat{a}_{\rm B}^\dagger|l\rangle|^2}{1+Q_{\rm
	B}^2(\frac{f_{kl}}{f_{{\rm r}}}-\frac{f_{{\rm
	r}}}{f_{kl}})^2}\frac{(f_{kl}/f_{\rm r})^2}{1-e^{-\beta_{\rm
	B}hf_{kl}}}\, .
\end{equation}


\newpage
\begin{thebibliography} {99}
\bibitem{vinjanampathy_quantum_2016} Vinjanampathy, S. \& Anders, J. Quantum
	thermodynamics. Contemp. Phys. {\bf 57}, 545-579 (2016).

\bibitem{goold_role_2016} Goold, J. \& Huber, M. \& Riera, A. \& del Rio, L. \&
	Skrzypczyk, P. The role of quantum information in
		thermodynamics{\textemdash}a topical review. J. Phys. A: Math.
		Theor. {\bf 49}, 143001 (2016).

\bibitem{martinezheat} Mart\'inez-P\'erez, M. J. \& Giazotto, F. The Josephson
	heat interferometer, Nature {\bf 492}, 401-405 (2012).

\bibitem{pekola2015} Pekola, J. P. Towards Quantum Thermodynamics in electronic
	circuits. Nat. Phys. {\bf 11}, 118-123 (2015).

\bibitem{jezouin}Jezouin, S. \& Parmentier, F. D. \& Anthore, A. \& Gennser U.
	\& Cavanna, A. \& Jin, Y. \& Pierre, F. Quantum limit of heat flow
		across a single electronic channel. Science {\bf 342}, 601-604
		(2013).

\bibitem{schwab}Schwab, K. \& Henriksen, E. \& Worlock, J. \& Roukes, M.
	Measurement of the quantum of thermal conductance. Nature {\bf 404},
		974-977 (2000).

\bibitem{banerjee}Banerjee, M. \& Heiblum, M. \& Rosenblatt, A. \& Oreg, Y. \&
	Feldman, D. E. \& Stern, A. \&  Umansky, V. Observed quantization of
		anyonic heat flow. Nature {\bf 545}, 75-79 (2017).

\bibitem{sivre}Sivre, E. \& Anthore, A. \& Parmentier, F. D. \& Cavanna, A. \&
	Gennser, U. \& Ouerghi, A. \& Jin, Y. \& Pierre, F. Heat Coulomb
		blockade of one ballistic channel. Nat. Phys. {\bf 14}, 145-148 (2018).

	\bibitem{cottet} Cottet, N. \& Jezouin, S. \& Bretheau, L. \&
		Campagne-Ibarcq, P. \& Ficheux, Q. \& Anders, J. \& Auff\`eves,
		A. \& Azouit, R. \& Rouchon, P. \& Huard, B. 
		Observing a quantum Maxwell demon at work.
	Proc. Nat. Acad. Sci. {\bf 114}, 7561-7564 (2017).

\bibitem{partanen2017} Partanen, M. \& Tan, K. Y. \& Masuda, S. \&
	Govenius, J. \& Lake, R. E. \& Jenei, M. \& Gr\"onberg, L. \& Hassel, J.
		\& Simbierowicz, S. \& Vesterinen, V. \& Tuorila, J. \& 
		Ala-Nissila, T. \& M\"ott\"onen, M.
		Flux-tunable heat sink for quantum electric circuits.
		Sci. Rep. {\bf 8} 6325 (2018).

\bibitem{wallraff} Wallraff, A. \& Schuster, D. I. \& Blais, A. \& Frunzio, L.
	\& Huang, R. -S. \& Majer, J. \& Kumar, S. \& Girvin, S. M. \&
		Schoelkopf, R. J. Strong coupling of a single photon to a
		superconducting qubit using circuit quantum electrodynamics.
		Nature {\bf 431}, 162-167 (2004).

\bibitem{neill_ergodic_2016} Neill, C. \& Roushan, P. \& Fang, M. \&
	Chen, Y. \& Kolodrubetz, M. \& Chen, Z. \& Megrant, A. \& Barends, R. \&
		Campbell, B. \& Chiaro, B. \& Dunsworth, A. \& Jeffrey, E. \&
		Kelly, J. \& Mutus, J. \& O'Malley, P.J.J. \& Quintana, C. \&
		Sank, D. \& Vainsencher, A. \& Wenner, J. \& White, T.C. \&
		Polkovnikov, A. \& Martinis, J.M.
	Ergodic dynamics and thermalization in an isolated quantum system. Nat.
		Phys. {\bf 12}, 1037-1041 (2016).

\bibitem{srednicki_chaos_1994} Srednicki, M. Chaos and quantum thermalization.
	Phys. Rev. E {\bf 50}, 888 (1994).

\bibitem{kaufman} Kaufman, A. M. \& Tai, M. E. \& Lukin, A. \& Rispoli, M. \& Schittko,
	R. \& Preiss, P. M. \& Greiner, M. Quantum thermalization through
		entanglement in an isolated many-body system. Science {\bf 353},
		794-800 (2016).

\bibitem{reimann_eigenstate_2015} Reimann, P. Eigenstate thermalization:
	{Deutsch}{\textquoteright}s approach and beyond. New J. Phys. {\bf 17},
		055025 (2015).

	\bibitem{koch_charge-insensitive_2007} Koch, J. \& Yu, T. M. \&
		Gambetta, J. \& Houck, A. A. \& Schuster, D. I. \& Majer, J. \&
		Blais, A. \& Devoret, M. H. \& Girvin, S. M. \& Schoelkopf, R.
		J. Charge-insensitive qubit design derived from the {Cooper}
		pair box. Phys. Rev. A {\bf 76}, 042319 (2007).

\bibitem{rossnagel_single_2016} Ro{\ss}nagel, J. \& Dawkins, S. T. \& Tolazzi,
	K. N.  \& Abah, O. \& Lutz, E. \& Schmidt-Kaler, F. \& Singer, K. A
	sigle atom heat engine. Science {\bf 352}, 325-329 (2016).

\bibitem{kosloff}Kosloff, R. \& Levy, A. Quantum heat engines and refrigerators:
	continuous devices. Annual Rev. Phys. Chem. {\bf 65}, 365-393 (2014).

\bibitem{karimi_otto_2016} Karimi, B., \& Pekola, J. P. Otto refrigerator based
	on a superconducting qubit: Classical and quantum performance. Phys.
	Rev. B {\bf 94}, 184503 (2016).

\bibitem{schmidt_photon-mediated_2004} Schmidt, D. R. \& Schoelkopf, R. J. \&
	Cleland, A. N. Photon-mediated thermal relaxation of electrons in
	nanostructures. Phys. Rev. Lett. {\bf 93}, 045901 (2004).

\bibitem{meschke} Meschke, M. \& Guichard, W. \& Pekola, J. P. Single-mode heat
	conduction by photons. Nature {\bf 444}, 187-190 (2006).

\bibitem{partanen2016}Partanen, M. \& Tan, K. Y. \& Govenius, J. \& Lake, R. E.
	\& M{\"a}kel{\"a}, M. K. \& Tanttu, T. \& M{\"o}tt{\"o}nen, M.
	Quantum-limited heat conduction over macroscopic distances. Nat. Phys.
	{\bf 12}, 460-464 (2016).
	
\bibitem{pothier} Pothier, H. \& Gueron, S. \& Birge, N. O. \& Esteve, D. \& Devoret,
	M. H. Energy distribution function of quasiparticles in mesoscopic
	wires. Phys. Rev. Lett. {\bf 79}, 3490 (1997).

\bibitem{simone} Gasparinetti, S. \& Viisanen, K. L. \& Saira, O.-P. \& Faivre,
	T. \& Arzeo, M. \& Meschke, M. \& Pekola, J. P. Fast electron thermometry
	towards ultra-sensitive calorimetric detection. Phys. Rev. Appl. {\bf
	3}, 014007 (2015).

\bibitem{govenius_microwave_2014}Govenius, J. \& Lake, R. E. \& Tan, K. Y. \&
	Pietil{\"a}, V. \& Julin, J. K. \& Maasilta, I. J. \& Virtanen, P., \&
	M{\"o}tt{\"o}nen, M. Microwave nanobolometer based on proximity
	Josephson junctions. Phys. Rev. B {\bf 90}, 064505 (2014).

\bibitem{tan_quantum-circuit_2017} Tan, K. Y. \& Partanen, M. \& Lake, R. E. \&
	Govenius, J. \& Masuda, S. \& M{\"o}tt{\"o}nen, M. Quantum-circuit
	refrigerator. Nat. Comm. {\bf 8}, 15189 (2017).
	
\bibitem{revmodgiaz} Giazotto, F. \& Heikkilä, T. T. \& Luukanen, A. \& Savin,
	A. M. \& Pekola, J. P. Opportunities for mesoscopics in thermometry and
	refrigeration: Physics and applications. Rev. Mod. Phys {\bf 78}(1)
	217-274 (2006).

\bibitem{pendry_quantum_1983} Pendry, J. B. Quantum limits to the flow of
	information and entropy. J. Phys. A: Math. Gen. {\bf 16}, 2161 (1983).

\bibitem{bianchetti} Bianchetti, A. R. Control and readout of a superconducting
	artificial atom. Doctoral Thesis, ETH Zurich (2010).

\bibitem{Jukka1} B. Karimi, J. P. Pekola, Otto refrigerator based on a superconducting qubit: Classical and quantum performance, Phys. Rev. B {\bf 94}, 184503 (2016).

\bibitem{Jukka2} B. Karimi, J. P. Pekola, M. Campisi, and R. Fazio, Coupled qubits as a quantum heat switch, Quantum Sci. Technol. {\bf 2}, 044007 (2017).

\end{thebibliography}
\end{document}